\documentclass[conference]{IEEEtran}
\usepackage{graphicx}
\usepackage{subfigure}
\usepackage{times}
\graphicspath{{images/}}

\usepackage{epsf}
\usepackage{url}
\usepackage{algorithm}
\usepackage{algorithmic}

\author{Mohammed Ali Kaafar, Thierry Turletti, Walid Dabbous\\
INRIA Sophia Antipolis, France \\
E-mail:{\{mkaafar, turletti, dabbous\}@sophia.inria.fr}}

\title{A Locating-First Approach for Scalable Overlay Multicast}

\begin{document}

\maketitle

\begin{abstract}
Recent proposals in multicast overlay construction have
demonstrated the importance of exploiting underlying network
topology. However, these topology-aware proposals often rely on
incremental and periodic refinements to improve the system
performance. These approaches are therefore neither scalable, as
they induce high communication cost due to refinement overhead,
nor efficient because long convergence time is necessary to obtain
a stabilized structure. In this paper, we propose a highly
scalable locating algorithm that gradually directs newcomers to
their a set of their closest nodes without inducing high overhead.
On the basis of this locating process, we build a robust and
scalable topology-aware clustered hierarchical overlay scheme,
called LCC. We conducted both simulations and PlanetLab
experiments to evaluate the performance of LCC. Results show that
the locating process entails modest resources in terms of time and
bandwidth. Moreover, LCC demonstrates promising performance to
support large scale multicast applications.
\end{abstract}

%

\section{Introduction}

A key factor to ``overlay networking'' success is that an overlay
service can be quickly constructed and easily deployed and
upgraded. In particular, several overlays support multicast-style
data dissemination service without requiring the widespread
deployment of IP multicast. However, such application-level
multicast may suffer from poor performance, scale and cost
problems when the delivery tree construction process ignores the
topology and link characteristics of the underlying network. If an
overlay is built in this way, nearby nodes in the overlay network
may actually be distant from each others in the underlying
network. Recent proposals in multicast overlay construction [1-10]
demonstrate the importance of exploiting underlying network
topology. However, we claim that there are barriers for quality of
service aspects, namely scalability and efficiency in existing
topology-aware overlay multicast protocols:

\begin{enumerate}
    \item Although decentralized protocols have been designed to be scalable, by not relying on global
    knowledge, they often rely on periodic and incremental refinement processes, which
induce high overhead. In these protocols such
as~\cite{ESM}~\cite{swithctrees}~\cite{Hostcast}~\cite{MeshTree},
nodes maintain their relative positions to the root of the
delivery tree. Periodically, each node tries to improve its
on-tree position by finding a better parent, i.e. a non-descendant
node that provides a lower delay to the root. Therefore, these
protocols are generally not scalable to support large multicast
groups. Additional overhead is incurred in case of dynamicity of
either the overlay membership or the underlying network
conditions. In fact, during overlay growth or membership changes,
heavy control overhead is incurred due to periodical structure
quality maintenance and partition repairs operations. On the other
hand, higher frequency control messages is required to map the
overlay to varying network characteristics.

    \item Users attending a video conferencing session or
an event broadcast expect an acceptable quality as soon as they
join the multicast session. Since a multicast overlay delivery
tree is typically studied to minimize the average incurred delay
observed by the receivers, we consider that a delivery tree is
``efficient'' if the average incurred delay is less than a
threshold value. However, one would expect that incremental
refinements-based approaches incur a long delay before the overlay
delivery tree converges to an efficient structure.
\end{enumerate}

In this paper, we provide a practical solution for large-scale and
efficient multicast support. First, we propose a simple and
accurate location-aware process for connecting new members to the
overlay network. The basic idea is to use the nodes in the already
constructed overlay to suggest candidate neighbors that are close
to a newcomer. The latter gradually requests the suggested nodes
to refine its localization in the underlying network. This
locating process does not use virtual coordinates system embedding
nor fixed landmarks measurements, and aims to be accurate and
scalable.

Second, we build a robust and scalable topology-aware clustered
hierarchical overlay on the basis of the locating process. We
propose proactive mechanisms to react to cluster leaders failures,
and to smoothly manage overlay topology changes caused by crash
scenarios or underlying network changes. Scalability is achieved
by drastically reducing the bandwidth requirements for overlay
maintenance. Robustness is obtained by mitigating the effect of
dynamic environment as most changes are quickly recovered and not
seen beyond clustered set of nodes. The proposed overlay multicast
construction scheme, called Locate, Cluster and Conquer (LCC), has
been designed to address the aforementioned quality of service
issues. Intuitively, running the locating process before that the
node joins the overlay, and then clustering nearby nodes should
allow to perform fast convergence to an efficient multicast
delivery tree. Furthermore, it would reduce management overhead
and delivery tree changes imposed due to periodical refinements.
However, these enhancements could be mitigated by the overhead of
the locating process.

Taking into account these considerations, we evaluated the LCC
scheme using two complementary evaluation methods: simulations and
experimentations over the PlanetLab testbed. Results show that LCC
has low overhead upon the locating process and during the session.
Compared to other initially-randomly and topology-aware
approaches, LCC achieves lower convergence time and performs less
link adjustments rate. At the same time, it performs well in terms
of data distribution efficiency even in large overlays.


The remainder of this paper is structured as follows. Section 2
presents the related work. Section 3 provides an overview of the
LCC scheme. The locating process is detailed in Section 4. Then
the clustering mechanism and its different components are
presented in Section 5. Experiments and simulations are discussed
in Section 6 and a comparison with various previous approaches is
provided. Finally, Section 7 concludes the paper.

\section{Related Work}

There has been tremendous interest in the construction of overlays
to provide application-level multicast. Basically, the
contributions can be categorized in two classes: overlay-router
approach and P2P approach.

In the overlay-router approach such as OMNI~\cite{OMNI} and
TOMA~\cite{LA05}, reliable servers are installed across the
network to act as application-level multicast routers. The content
is transmitted from the source to a set of receivers on a
multicast tree consisting of the overlay servers. This approach is
designed to be scalable since the receivers get the content from
the application-level routers, thus alleviating bandwidth demand
at the source. However, it needs dedicated infrastructure
deployment and costly servers.

The P2P approach requires no extra resources. Several proposals
have been designed to handle small groups. Narada~\cite{ESM},
MeshTree~\cite{MeshTree}, and Hostcast~\cite{Hostcast} are
examples of distributed ``mesh-first'' algorithms where nodes
arrange themselves into well-connected mesh on top of which a
routing protocol is run to derive a delivery tree. These protocols
rely on incremental improvements over time by adding and removing
mesh links based on an utility function. Although these protocols
offer robustness properties (thanks to the mesh structure), they
do not scale to large population, due to excessive overhead
resulting from the improvement process. The objective of LCC is to
locate the newcomer prior to joining the overlay and hence process
only a few number of refinements during the multicast session.

Other ``tree-first'' protocols like ZigZag~\cite{ZigZag} and
NICE~\cite{NICE}, are topology-aware clustering-based protocols
which are designed to support wide-area size multicast for low
bandwidth application. However, they do not consider individual
node fan-out capability. Rather, they bound the overlay fan-out
using a (global) cluster-size parameter. In particular, since both
protocols only consider latency for cluster leader selection, they
may experience problems if the cluster leader has insufficient
fan-out. Other proposals exploit the AS-level~\cite{S004} or the
router-level~\cite{KW02} underlying network topology information
to build efficient overlay networks. However, these approaches
assume some assistance from the IP layer (routers sending ICMP
messages, or BGP information access), which may be problematic.
LCC does not require any extra assistance from entities that do
not belong to the overlay.

Landmark clustering is a general concept to construct
topology-aware overlays. Ratnasamy et al.~\cite{Top-CAN} use such
an approach to build a multicast topology-aware CAN overlay
network. Prior to joining the overlay network, a newcomer has to
measure its distance to each landmark. The node then orders the
landmarks according to its distance measurements. The main
intuition is that nodes with the same landmark ordering, are also
quite likely to be close to each other topologically. An immediate
issue with such a landmark-based approach is that it can be rather
coarse-grained depending on the number of landmarks used and their
distribution. Furthermore, requiring a fixed set of landmarks
known by all participating nodes renders this approach unsuitable
for dynamic networks.

\section{Overview of LCC}

We have designed a two-level clustered overlay multicast
architecture (LCC) to provide scalable, efficient and robust
multicast distribution service to end users. Basically, the LCC
overlay construction is divided into two processes:
\emph{Locating} and \emph{Clustering}.

The \textbf{locating process} aims to direct newcomers to the
``nearest'' cluster before they receive data on the delivery tree.
A newcomer initiates the locating process by sending a
``Localization\_Request'' to a randomly selected cluster leader
(denoted by boot node). According to its location-information
knowledge, the boot node selects a few cluster leaders (that we
will denote \emph{the queried nodes}) that it considers to be
close to the newcomer. It asks them to probe the newcomer, and
gets each queried node's answer. Then, it suggests to the newcomer
the possible closest nodes. By iteratively sending
``Localization\_Request'' messages to the closest nodes (called
\emph{the requested nodes} in the rest of the paper), the newcomer
is able to gradually locate nodes that are close by. Each
requested node uses a selection criterion to limit the number of
nodes probing the newcomer, hence minimizing the locating
overhead. The locating process ends by proposing one or more
nearby cluster leaders.

By grouping together nodes that are close to a cluster leader,
members are expected to be close to each other, which leads to low
overhead of intra-cluster control messages. The \textbf{clustering
process} is initiated by every node once the locating process
terminates. On the basis of their locating result, nodes are
partitioned into clusters of nodes. A maximum distance, $R_{max}$,
defines the interval in which other nodes are considered
``nearby''. This interval is called the cluster leader's
\emph{scope}, and defines the clustering criterion. During the
clustering process, a node decides at which level it will join the
overlay. If it creates its own cluster, it joins the ``top-level''
topology and starts an inter-cluster mesh construction. Otherwise,
it becomes a cluster member and joins an intra-cluster mesh in
order to derive its delivery tree.

\begin{figure} [htb] \centering
\includegraphics[width=3.5in]{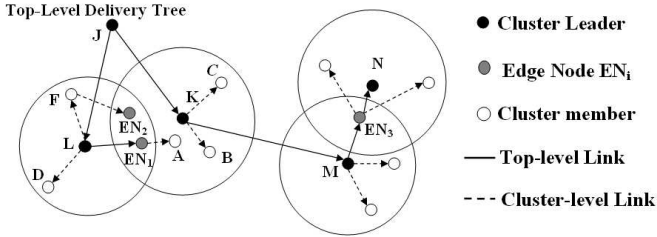}
\caption{The two-level hierarchy of LCC.} \label{clusters}
\end{figure}

Since a node could be in more than one cluster leader's scope, it
could be member of more than one cluster. Such nodes are called
\emph{edge nodes}. We exploit edge nodes to improve the overlay
efficiency. In fact, the cluster leader is the primary responsible
of connecting its cluster to the top-level overlay. Nevertheless,
edge nodes are also allowed to join the inter-cluster mesh at the
top level. The main role of edge nodes is to allow (if fan-out
constraints are not violated) the clusters members to derive their
delivery tree considering the edge node as an alternative nearby
source connected to the top-level topology. Moreover, these nodes
may contribute in the overlay robustness in case of cluster
leaders failures. Although edge nodes are attached to several
meshes of different clusters, they do not receive the data several
times. In fact, as each edge node derives a unique delivery tree
from one of the existing intra-cluster mesh, it is then a child in
this particular delivery tree. On the other hand, it could be a
parent in several derived delivery trees in other clusters. A high
level picture of LCC is illustrated in Fig.~\ref{clusters}.

Note that LCC does not specify a new tree construction protocol;
any existing protocol may be used on top of LCC. In this paper, we
choose to construct the LCC overlay by running the MeshTree
protocol~\cite{MeshTree} at both the top-level and the
intra-cluster level. MeshTree embeds the delivery tree in a
degree-bounded mesh containing many low-cost links. The
constructed mesh consists then of two main components: (i) a
backbone structure, composed of a low-cost tree and connecting
nodes that are topologically close together, and (ii) additional
links to improve the delay properties. The delivery tree is then
derived from the mesh using a path-vector routing protocol. The
``Flat'' MeshTree first constructs a randomly connected overlay
and relies on periodical adding/deleting links using a set of
local rules. Unlike this approach, the LCC scheme, initially
constructs location-aware overlay based on the locating and
clustering processes. Top-level nodes then act as particular
MeshTree nodes, where other clusters represent neighbors in the
derived delivery tree (see Fig.~\ref{clusters}).

In order to construct an overlay spanning tree rooted at the
source node $s$, we need to consider the degree constraints.
Assuming that the media playback rate is $R$ and the outgoing
access link capacity of any particular node $i$, is $c_{i}$, the
total number of streams that the node can handle is
$\emph{f}_{i}^{max}$ = [$c_{i}/R$]. The fan-out value of node $i$
represents the maximum number of connections that this node can
establish with other nodes. We assume that each node can estimate
its connection type (eg ADSL, 802.11, etc.), relying on system and
user specifications. Moreover, LCC can use a history of maximum
throughput of the most recent downloads, as an indication of its
effective connection speed. These fan-out estimation techniques
are used in order to avoid each node to measure its available
bandwidth, which may involve high overhead. We also define the
cluster overall capacity as $\sum_{i=0}^{m}f_{i}^{max}-m$, where
$m$ is the number of members in the clusters. Next, we detail both
the locating and the clustering LCC processes.

\section {The Locating Process}

\begin{figure} [htb] \centering
\includegraphics [width=3.5in]{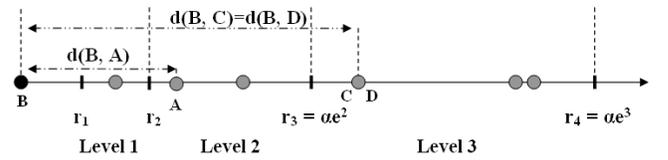}
\caption{The locating system of node B.} \label{locating}
\end{figure}

LCC adopts a nodes' positioning strategy similar to
meridian~\cite{Meridian} to organize nodes into levels according
to a distance metric. Typically, the distance between two nodes is
the round trip network delay. Each LCC node keeps track of a fixed
number of other nodes in its locating system. A locating system is
a set of non overlapping and exponentially increasing levels,
represented by intervals $[r_{i},r_{i+1}[$, where $r_i = \alpha
e^{i-1}$ for $i\geq1$ and $r_{0} = 0$ (see Fig.~\ref{locating}).
 Nodes measure the distances to the set of
nodes they are aware of, and assign each node a position in the
correspondent level of their locating system. For example if the
measured distance $d$ satisfies $r_i\leq d< r_{i+1}$, the node is
positioned in the $i^{th}$ level. All considered nodes in the
locating system are cluster leaders. In the following, we describe
the locating process operations.

\subsection{Bootstrap and locating request}

Initially, a newcomer, say node $A$, has to contact a global
well-known Rendezvous Point\footnote{or any equivalent mechanism.}
to obtain the identity of a randomly selected boot node, $B$. Node
$A$ measures the distance from itself to $B$, $d(A,B)$ and assigns
$B$ a level in its locating system, say level $i$. If $A$ is in
$B$'s scope, i.e. $d(A,B) \leq R_{max}$, the clustering criterion
is met and the locating process terminates, and $A$ sends a
request to join $B$'s cluster. Otherwise, $A$ sends $B$ a
``Localization\_Request''. Upon receiving such request, the
requested node $B$ simultaneously queries nodes that it considers
as nearby to $A$. These queried nodes have then to report the
results back to the requested node. If a queried node is closer to
the newcomer than the requested node, it is considered as a
candidate. A list that identifies the set of candidate nodes is
sent by the requested node to the newcomer $A$. Among this list,
$A$ initiates cluster joining processes with all nodes that meet
the clustering criterion. If there are no such nodes in the list,
nodes in the candidate list become possible requested nodes, since
$A$ re-initiates the locating process with each node in this list
sorted in increasing distances. The list is updated at each
response from a requested node. This procedure is repeated until
the newcomer finds a cluster leader in its scope. Finally, it is
necessary to set a stop criterion to terminate the process within
a given time period by repeating the procedure at most $C$ times.
If the algorithm ends without satisfying the clustering criterion,
$A$ creates its own cluster.

\subsection{The selective-locating process}

During the locating request, each requested node has to query a
set of nodes. It then selects among them a list of candidate nodes
to send to the newcomer. In this subsection, we answer the
following question: How queried nodes are chosen by the requested
node?

The basic solution would be that the requested node asks all the
nodes in the same level than the newcomer and in the adjacent
levels, as potential queried nodes. To establish a reference, we
consider this solution that we call the ``non-selective'' locating
process. Although it has the advantage of simplicity, this
solution may induce high overhead. In fact, while being in the
same or adjacent levels than the newcomer, some queried nodes
should not be taken into consideration for probing the newcomer,
since they may be not closer to it than the requested node.

We introduce the \emph{selection criterion} in order to reduce the
number of useless probes during the locating process. Basically,
the ``selective locating'' consists in querying only specific
representative nodes. Nodes that are close enough to a
representative node, randomly selected by the requested node, are
not queried to measure their distance to $A$: the less queried
nodes, the less measurements and control overhead.

Closeness is defined by a distance threshold value $\gamma_{i}$,
which is a function of the distance between the newcomer and the
requested node, $d$. If the newcomer is close to the level
frontier or to the requested node, the latter should use a
fine-grained selection and a small $\gamma_{i}$ value should be
used. If not, the requested node should use a greater $\gamma_{i}$
value. In our algorithm, we choose:
\begin{displaymath} \gamma_{i} = \frac{|d-r_{i}|}{r_{i+1}}\times d \end{displaymath}

Nodes maintain for each level $i$ a square matrix, $M^{i}$,
representing learned distances of level $i$'s nodes to each other,
and to nodes in adjacent levels $i-1$ and $i+1$. Values in $M^{i}$
are assigned as and when discovered through other nodes' locating
requests. If a distance is not known, it is set to a value large
enough to discard the concerned node from the selection. Each
element $M^i(j,k) = d(N^i_j, N^i_k)$ in $M^{i}$ corresponds to the
distance between nodes $N^i_j$ and $N^i_k$. The $j^{th}$ row in
$M_i$ represents the learned distances between node $N^i_j$ and
other nodes in level $i$ and adjacent levels. The selection
algorithm run by a requested node is presented in
Algorithm~\ref{algoSelection}
\begin{algorithm} \caption{Selection} \label{algoSelection}

\begin{algorithmic} [h]
{\footnotesize
\REQUIRE Distance
\ENSURE List of representative nodes to query
\STATE $Level \leftarrow Assign\_Level(Distance)$
\STATE $Candidates \leftarrow Search\_Nodes (Level)$
\STATE $S^{Level} \leftarrow Get\_Distance\_Matrix (Candidates)$
\STATE $Threshold \leftarrow \frac{|Distance - r_{Level}| \times
Distance}  {(r_{Level+1})}$

\REPEAT

\STATE $j \leftarrow Random(Dimension(S^{Level}))$

\STATE $V \leftarrow Extract\_Row (S^{Level}, j)$

\FOR{$i\in Dimension(S^{Level})$}

\IF{$V(i)<Threshold$}

\STATE $Represented \leftarrow Represented \cup
{Index\_to\_Node(i)}$
\ENDIF

\ENDFOR

\STATE $Representative \leftarrow Representative +
{Index\_to\_Node(j) }$

\STATE $ S^{Level} \leftarrow S^{Level}\setminus
Columns(Represented)$

\UNTIL {$Elements(S^{Level}) = Representative$}

\STATE Return $Representative$
}
\end{algorithmic}
\end{algorithm} and can be described as follows:
Each requested node selects a random node, $N^i_j$, from level $i$
or adjacent levels. If $M^i(j,k) = d(N^i_j, N^i_k)$ is less than
the threshold value $\gamma_{i}$, then node $N^i_k$ is represented
by $N^i_j$. Selected nodes are represented by a matrix, say $S^i$,
which is initially equal to $M^i$. At each iteration of the
selection process, $S^i$ is diminished by the columns of nodes in
$M^i$ that can be represented by the selected node $N^i_j$. The
selection algorithm terminates when $S^i$ contains only distances
of representative nodes.

\section{The Clustering Process}

In this section, we describe the protocol to form and maintain
clusters. In this work, we emphasize mechanisms to enhance the
overlay QoS by increasing scalability and robustness. In
particular, we propose a proactive algorithm to manage failures of
leaders, and new cluster formation afterwards. We also propose new
mechanisms to smoothly manage cluster topology changes due to
leadership or underlying network changes.

\subsection{Cluster Creation}

In early stage of the overlay formation, new clusters are more
frequently generated since few nodes exist. If the locating
process ends with no leaders found in the newcomer's scope, the
latter creates its proper cluster with a new cluster ID. It then
contacts the closest cluster leaders that the locating process
returns, to join the top-level topology. Contacted cluster leaders
inform their members by flooding a ``New\_Cluster'' message.
Finally, members verify if they are in the new leader's scope,
i.e. if they are potential edge nodes.

\subsection{Cluster Joining}

A classical joining operation is initiated by a newcomer detecting
cluster leaders in its scope after the locating process
terminates. The newcomer sends simultaneously a ``Join\_Request''
message to all the detected cluster leaders. The request contains
its fan-out value and the set of other clusters it may belong to.
Upon receiving ``Join\_Notification'' messages, it sends
acknowledgement messages mentioning successfully joined clusters.

In LCC, each top-level node has two types of neighbors: nodes in
its own cluster and other top-level nodes. Since a cluster leader
has limited available bandwidth, it should carefully set its node
degree to maintain a balance between connecting to other top-level
nodes for better overall performance and serving as many nodes as
possible in its own cluster. If the cluster overall capacity is
$\geq$ 1, the cluster leader accepts the newcomer. Note that the
cluster overall capacity is null if all nodes are edge nodes
attached to the top-level topology. So, considering the case where
all of these nodes have a fan-out value of 1, this would lead to a
saturated cluster. This situation can be recovered if the cluster
leader requests an edge node to leave other cluster membership to
serve a newcomer.

If the newcomer is accepted , the cluster leader randomly assigns
it a cluster member to boot into the cluster-level mesh. The
newcomer gets cluster maintenance information from the cluster
members.

\subsection{Cluster's member state and Information updating}

Using ``Keep\_Alive'' messages exchanged by cluster members allows
to share cluster state, and to update cluster information.
Information about other overlay nodes is obtained using a simple
gossip style node discovery technique. Basically, a node, $i$
maintains a list of known nodes in the overlay. Periodically, $i$
randomly picks a node from the list, say $j$ and sends to it a
randomly-constructed set of other known members. Node $j$ updates
its own known nodes list and replies using the same procedure.
This simple informative exchange allows nodes to maintain a
minimal view of the overlay membership. Next, we discuss how this
knowledge affects the average locating process iterations.

\subsection{Leaders election}

The cluster leader election is based on the value of priority
vectors ($PV$) used to maintain a nodes' rank. Basically, a $PV$
is defined as: $ PV  =   <f^{max}, \frac{1}{DL}, T, \frac{1} {CD},
Migrated> $, where $\emph{DL}$ stands for the node's current
perceived Latency in the intra-cluster Delivery tree, $\emph{CL}$
denotes the minimum distance from the node to a known foreign
Cluster Leader, $\emph{T}$ represents how long the node has stayed
in the overlay, and $\emph{Migrated}$ is a boolean indicating if
the current cluster leader is included in the node's scope. The
priority value is computed as a linear combination of the first 4
components of $PV$ with decreasing weights. These priorities are
used to sort appropriate eligible nodes.

Nodes update periodically their $PV$. Each $PV$ is distributed as
part of ``Keep\_Alive'' messages. When receiving nodes' $PV$, a
node sorts the cluster members with increasing priorities. In
fact, cluster nodes construct a proactive rescue plan, where each
node maintains a \textbf{local cache} storing shared information.
The local cache consists in a sorted list of nodes that are
eligible to become cluster leaders. In a dynamic network
environment, a cluster leader may depart unexpectedly at any time.
If the leader fails, nodes will know it after a period of time as
they do no more receive the ``Keep\_Alive'' messages from the
leader. Meanwhile, the second node in the list becomes
automatically the leader and sends out ``Keep\_Alive'' messages.
If the second also fails, the third one will stand up, etc. It is
important to notice that for stability purposes, eligible nodes
that win the leader election at their joining process, are
maintained at a second position in the local cache, until their
life time in the cluster reaches a greater value.

\subsection{Dynamic Clusters topology}

In this subsection, we discuss the behavior of LCC in case of
cluster members migration outside their cluster due to new cluster
election or underlying topology changes. We distinguish three
different clustering states:

\begin{itemize}
    \item \textbf{Stabilized state}: where the cluster leader is included in the scope of each member of the cluster.
    \item \textbf{Temporary state}: where at least one node have migrated outside its original
    cluster.
    \item \textbf{Recovering state}: where during the temporary state, all migrated nodes know about other migrated nodes, and start to
    evolve towards a stabilized state.
\end{itemize}
We introduce an algorithm that allows the migrated nodes to
collaborate in order to create suitable new clusters after a
temporary state. It is based on the nodes' rank in the local
cache. It consists in a recursive procedure, where a potential
leader node asks subsequent nodes in the cache to join its
cluster, and triggers a recovering procedure for migrated nodes
that are not in its scope.

Basically, nodes verify at each local cache update operation
whether the current cluster leader is in their scope or not. If
not, they mark it as foreign cluster neighbor in their $PV$ with
$Migrated$ = 1. At each received $PV$, a migrated node updates a
set of other migrated nodes. The first ranked migrated node
initiates the process by creating a new cluster and by sending a
``Recovering\_Request'' to the next migrated node. The request
contains identities of other nodes that are already in the node's
scope. Hence, each node is able, through previous received
requests, to determine migrated nodes that can still be leaders
(eligible nodes). If the node is included in the requesting node's
scope, it sends a positive \emph{ACK} to join its cluster and
returns to a stabilized state. A node which sends a negative ACK,
verifies at each request if it has been contacted by all prior
ranked eligible nodes in the cache. In this case, it becomes a
cluster leader and initiates its proper recovering procedure by
sending requests to next nodes in the local cache. The recovering
algorithm terminates when contacting the last ranked migrated
node. It then informs its ``new'' cluster neighbors along with its
previous cluster leader of the cluster split. Finally, it switches
to a stabilized state and connects to the top-level topology.

\section{Performance Evaluation}

To evaluate and test the LCC scheme, we carried out simulations
and PlanetLab~\cite{PlanetLab} experiments. While the goal of
simulation studies is to assess the effectiveness of proposed
techniques for large scale overlays, the PlanetLab experiments aim
to illustrate the system performance under particular real-world
environments.

\subsection{Simulations and Experimentations}

\subsubsection{Simulation Setup}

Using the BRITE Internet topology generator~\cite{BRITE}, we
simulated up to $10^4$ nodes networks. We used a node bandwidth
reference model based on the Gnutella peer bandwidth distribution
observed by Saroiu et al.~\cite{SA02}. We modeled the node join
using Poisson distribution with an average rate of 60 node joins
per simulation tick. The duration distributions were modeled with
an exponential distribution of 0.01 as parameter.

\subsubsection {Experimentations on PlanetLab}

We implemented LCC in a C library, and built wrappers for
well-known IP-multicast applications (vic/rat, vlc)\footnote{The
LCC source code is available in the public domain and can be
downloaded from~\cite{IMPLEM}.}. We tested the LCC overlay over a
set of 212 wide spread PlanetLab nodes. The set consists of 90
nodes in U.S, 90 nodes in Europe and 32 nodes in Asia. All
experiments were conducted over several days in October 2005. In
this paper, we discuss a representative set of experimental
results using ``planetlab1.cs.cornell.edu'' as the data source.
This source generates a 560 kbits/s (70 kB/s) data stream sent to
all the other group members. Nodes join the overlay at the average
rate of one every 2 seconds. We remove the top 20\% and bottom
20\% of the measurement results and take the average of the
remaining values. In practice, pings have been conducted using
ICMP ECHO messages, and we use 10 consecutive pings for latency
measurements\footnote{Shen shows that latency measurements with 10
pings are sufficiently accurate~\cite{KEN04}.}.

\subsubsection {Metrics}

We evaluate the LCC scheme in terms of (1) scalability, by
studying the control overhead during both data distribution and
overlay joining. We also observe the link adjustment frequency and
the locating process resources usage (time and number of nodes
needed to locate a newcomer); (2) efficiency, by measuring several
common metrics that characterize the multicast overlay. In
particular, we measure the Average Relative Delay Penalty ($ARDP$)
which is defined as the average ratio between the overlay delay
($d'$) and the shortest path delay in the underlying network ($d$)
from $s$ to all other nodes: $\frac{1}{N-1}
\sum_{i=1}^{N-1}\frac{d'(s,i)}{d(s,i)}$, where $N$ is the number
of nodes in the overlay. By considering that the delivery tree
``converges'' or becomes ``efficient'' when the $ARDP$ is less
than a threshold value (say 2), we study the convergence time.
Then we plot the $ARDP$ and the stress on the link (which
represents the number of copies of an identical packet sent over a
single link), varying the overlay size; (3) robustness, by
verifying the scheme robustness to leaders failures and its
ability to recover in case of crash scenarios; (4) locating
process accuracy, by experimenting newcomers' behavior during the
locating process and their ability to locate their closest node in
the underlying network, within a large overlay.

\subsubsection{Comparison}

In order to compare LCC to initially randomly-connected overlays
relying on periodic refinements, we experiment a variant of LCC,
disabling the locating process and setting the scope value to
zero, thus emulating MeshTree behavior. We call this variant Flat
MeshTree. In our simulations, we also compare LCC to two
previously proposed multicast overlay structures: OMNI~\cite{OMNI}
as an infrastructure-based approach and ZIGZAG~\cite{ZigZag} as a
topology-aware hierarchical approach.

\subsection{Performance Results}
In the following, we report both simulation and experimental
results.

\subsubsection{Control overhead}

\begin{figure} [htb] \centering
\includegraphics[width=3.4in]{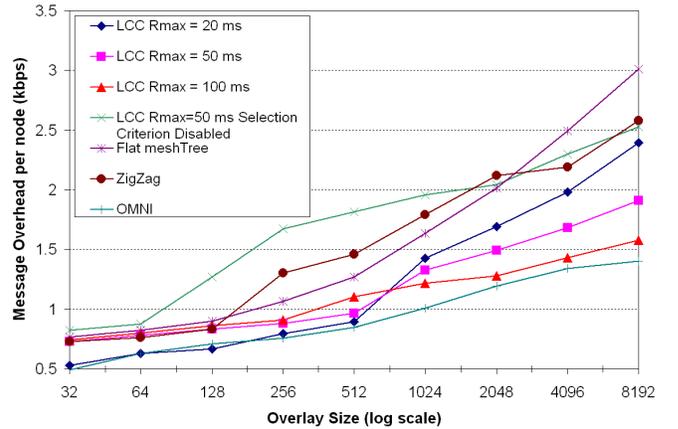}
\caption{Simulation of protocol overhead.}
\label{ProtocolOverhead}
\end{figure}

We ran simulations to evaluate the control traffic overhead in the
overlay and analyzed the protocol behavior in large size overlays.
We assumed a basic header size of 40 bytes per IP-packet and we
measured the overall control message traffic sent and received by
each node throughout a session. Fig.~\ref{ProtocolOverhead}

\begin{figure} [htb] \centering
\includegraphics[width=3.4in]{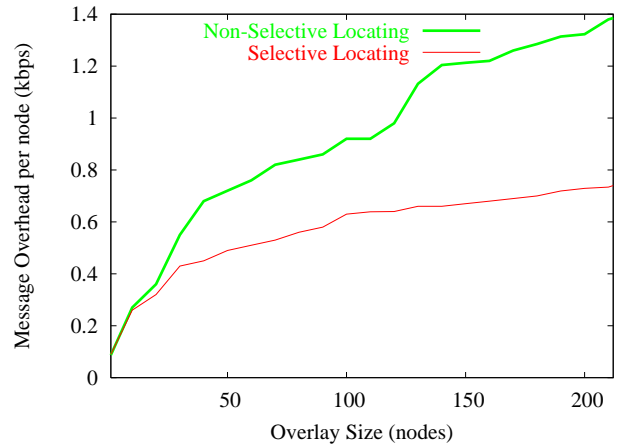}
\caption{PlanetLab: LCC overhead during the joining process.}
\label{overhead}
\end{figure} shows the average overhead per node when varying the overlay size.
Control overhead of LCC is lower than those of Flat MeshTree and
ZigZag, and is comparable to OMNI. We note that the per node
overhead in LCC, for different $R_{max}$ values, is steady for
small overlays. The maximum value reached for a 512 nodes overlay
is 1.10 kbps for LCC with $R_{max}$ = 100ms, and control messages
incur less than 2 kbps message overhead, in a 8000-nodes overlay.
OMNI nodes obtain lower control overhead. Since it is the
application-level multicast servers that are in charge of managing
the delivery tree, nodes in OMNI exchange a minimum number of
control messages to join the overlay. We note that the above
results include overheads due to network measurement, in
particular during the locating process, as we consider these
results from the instant the node contacts the Rendezvous Point.

To evaluate the cost of locating the closest cluster to join, we
measured on the PlanetLab testbed the average control traffic
overhead (in kbps) generated during overlay joining for both the
non-selective and selective locating process. We observe in
Fig.~\ref{overhead} the importance of the selection criterion
during the locating process. The per node overhead in the
selective locating process is reasonably small with about 0.7 Kbps
for a 212 nodes overlay. In addition, it increases very slowly
with the number of members. The locating messages are roughly 50\%
less frequent than those of a non-selective localization. Not
selecting nodes boosts the message overhead due to useless
measurement operations. In this case, requested nodes would
contact all the nodes in the newcomer's level and the adjacent
levels. These queried nodes will also measure their distance to
the newcomer, which would incrementally add network overhead.
However, we note that the selective locating process may require
more time to locate the newcomer. In fact, by selecting
representative nodes, the newcomer may need to contact more
requested nodes than the non-selective process as discussed later
and shown in Fig.~\ref{hopsKnownneighbors}.

\subsubsection{Link adjustment rate}

\begin{figure} [htb] \centering
\includegraphics[width=3.4in]{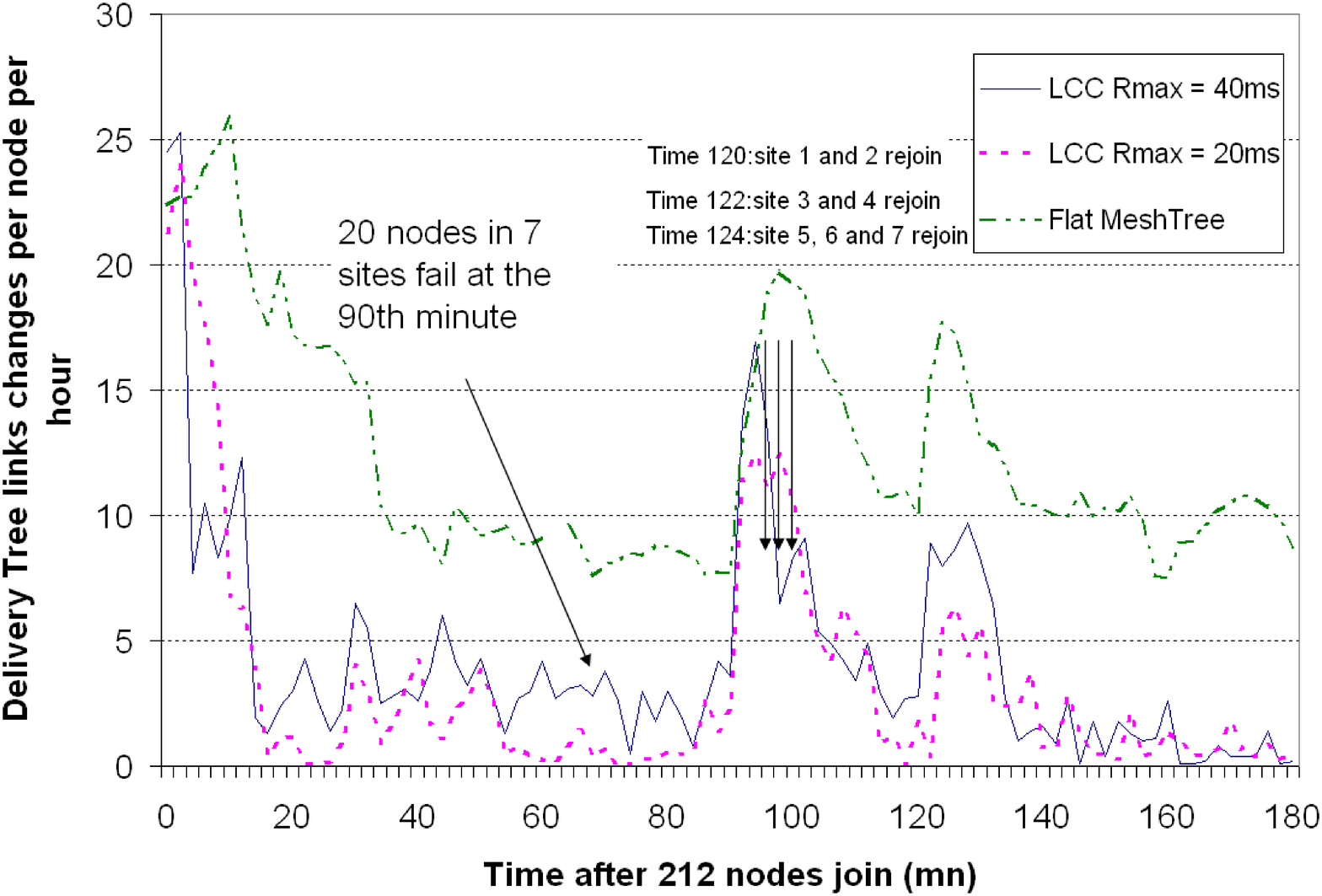}
\caption{PlanetLab: experimentation results of Link Adjustment
rate.} \label{LabAdjustement}
\end{figure}

Fig.~\ref{LabAdjustement} shows the LCC structure stability during
membership changes. On the PlanetLab testbed, we start tracking
the link adjustment counts right after the last node joined the
overlay. Results are collected at 30-second intervals. We observe
that the link adjustment rate mostly falls below 5 per hour per
node for the LCC overlay, whereas it stabilizes at roughly 10 per
node per hour for the Flat MeshTree. To confirm that the LCC
efficiency is robust, in case of membership frequent changes and
crash scenarios, we inject a simultaneous 20-nodes failure in 7
different sites at the $90^{th}$ minute and we let them rejoin the
overlay at the $120^{th}$ minute. We observe that the link
adjustment activity for LCC is moderate (mostly under 5 per hour
per node) during the membership changes. After the $140^{th}$
minute, the average link adjustment count falls around 2 per hour
per node. Due to randomness in initially connecting newcomers to
the clusters, the link adjustment rate of MeshTree is maintained
at 10. This assesses our intuition that non-initially locating
schemes may require high control messages for quality maintenance
and structure repairs operations.

\begin{figure} [htb] \centering
\includegraphics[width=3.4in]{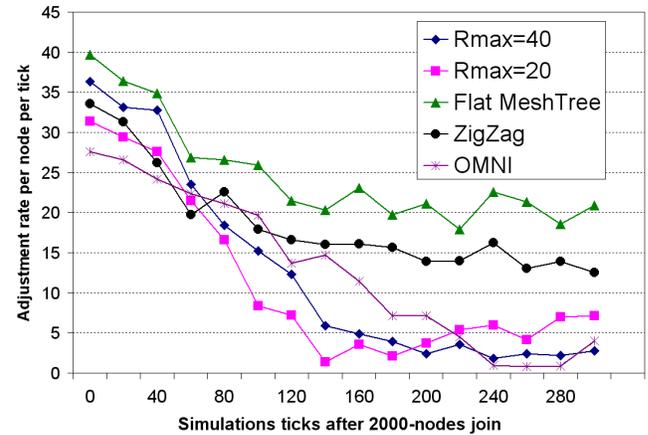}
\caption{Simulations: Link Adjustment rate results.}
\label{SimAdjustement}
\end{figure}

The simulation results shown in Fig.~\ref{SimAdjustement}

 confirm the PlanetLab experiments conclusions. Flat MeshTree
suffers from high adjustment rate, almost more than 20 links
change per node per simulation tick. Compared to ZigZag, the LCC
structure has a very low adjustment rate. This rate is stabilized
at less than 5 adjustments per node per tick. Link changes in OMNI
are less frequent than other improvement-based overlay structures.
OMNI nodes achieve an average link adjustment per node per tick of
12.9, with a minimum of 0.78. Nodes in LCC ($R_{max}$ = 20 ms)
have an average of 11.4 adjustments per node per tick with a
minimum links change of 1.4.

\subsubsection{Convergence Time}

\begin{figure} [htb]
\centering
\includegraphics[width=3.4in]{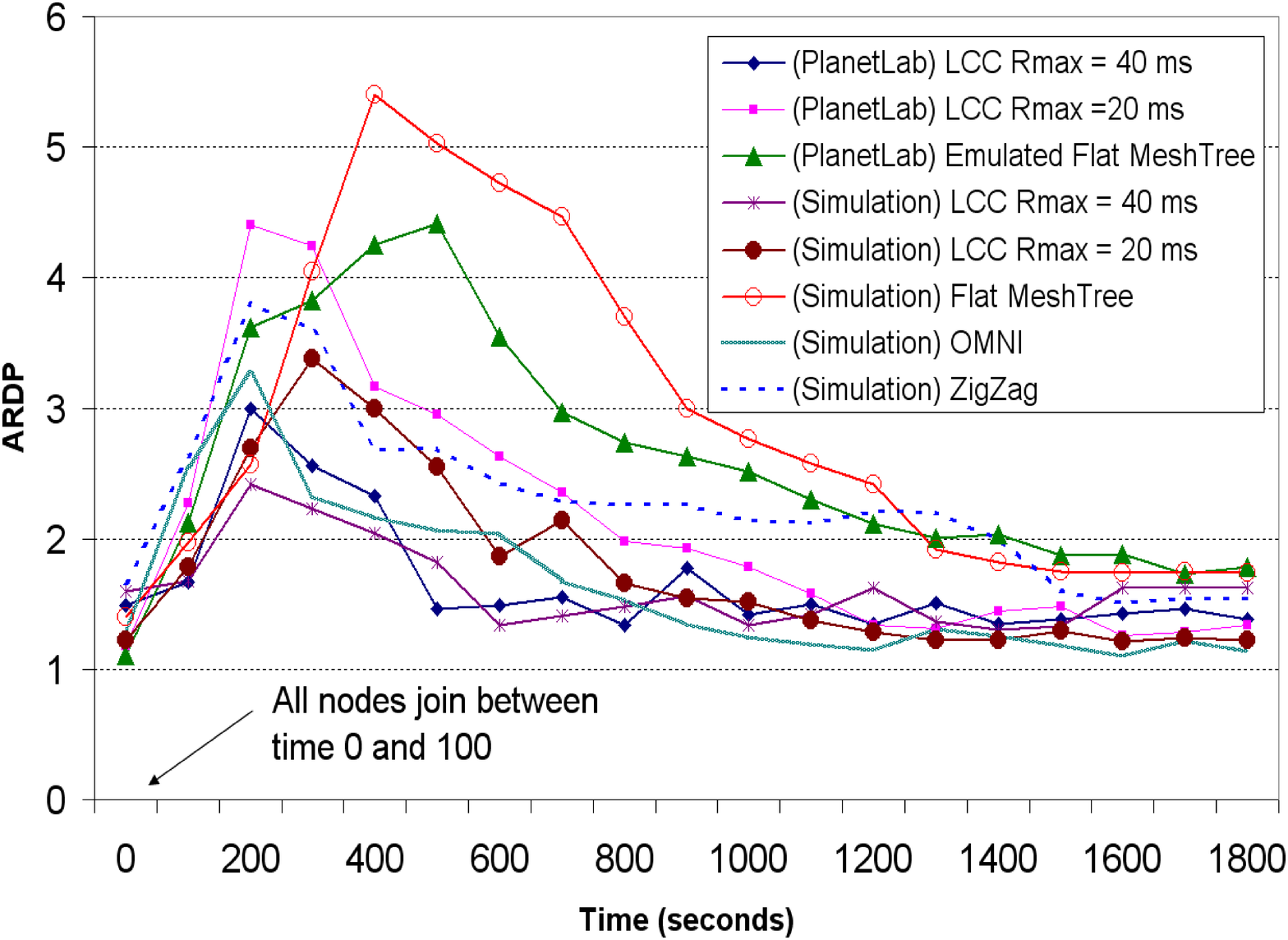}
\caption{Convergence Time property.} \label{convergence}
\end{figure}

The refinement-based approaches depend on the choice of a
refinement period, say $RT$. A small $RT$ value reduces the
convergence time, as more adjustment procedures are processed
within a short time, but may induce high overhead. A large $RT$
may reduce overhead at the expense of increased convergence time.
In the following experiments, we set the improvement period $RT$
to 30 seconds, for each receiver, and study the convergence time
metric which describes the overlay structure evolution in time.
Fig.~\ref{convergence} illustrates the convergence time, showing
$ARDP$ versus the multicast session time in both simulations
(Overlay size = 2000 nodes) and PlanetLab testbed. All nodes join
the overlay within the first 100 seconds. 

We observe that in LCC, $ARDP$ rapidly decreases to a value less
than 2 after the first 400 seconds, i.e. less than 14 improvement
rounds per node. For Flat MeshTree, it takes much more time for
$ARDP$ to stabilize with almost 1300 seconds. This shows the
importance of pro-actively organizing the overlay, to converge
very quickly to an efficient structure. The ZigZag overlay reaches
an acceptable $ARDP$ value much more quickly than MeshTree.
Although stabilized, this value is more than 2, which is still
inefficient to consider the overlay converged. The convergence
time of ZigZag is around 1400 seconds when $ARDP$ falls under 2.
The reason why ZigZag does not converge quickly is that during
overlay growth, several group merges and splits are imposed to not
exceed the maximal group size. This may induce several
improvements rounds, and link adjustments. The OMNI server-based
structure is not affected by frequent membership changes and
converges quickly, similarly to LCC.

\subsubsection{The Average incurred delay}

\begin{figure}
[htb] \centering
\includegraphics[width=3.4in]{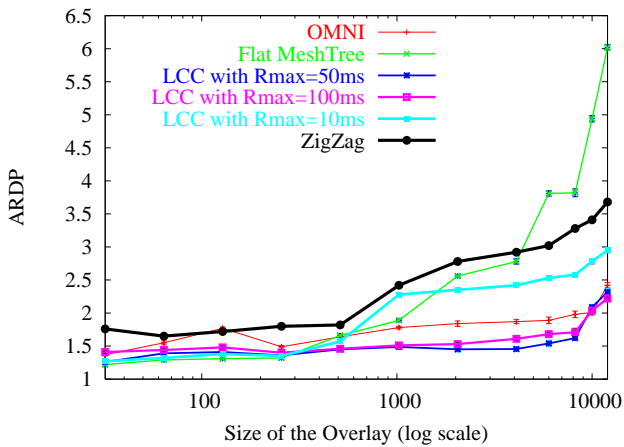}
\caption{Simulations: Average RDP property.}
\label{AverageRDPComparisonVaryingOverlaySize}
\end{figure}

We characterize the average incurred delay observed by the
receivers in a large populated overlay by observing the $ARDP$
variation according to the overlay size in
Fig.~\ref{AverageRDPComparisonVaryingOverlaySize}. 

\begin{figure} [htb] \centering
\includegraphics[width=3.4in]{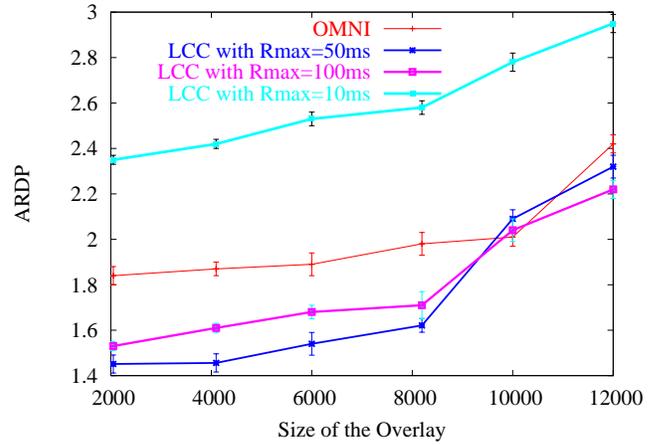}
\caption{Simulations: A zoomed in view of Average RDP variations.}
\label{ZoomAverageRDPComparisonVaryingOverlaySize}
\end{figure} In Flat MeshTree, the $ARDP$ increases drastically to more than 4
demonstrating that this protocol does not scale to a few thousands
of nodes. Nevertheless, we note that Flat MeshTree has lower
$ARDP$ than LCC structures in small groups (overlay size $\leq$
500 nodes). ZigZag maintains a stable $ARDP$ value while the
overlay size is increasing but suffers relatively poor performance
with $ARDP \geq$ 2.5 in a 3000-nodes overlay. To make it easier to
read, we zoom in on a portion of the graph in
Fig.~\ref{ZoomAverageRDPComparisonVaryingOverlaySize}. We observe
that the $ARDP$ of LCC is about 60\% of ZigZag. For $R_{max}$=50
ms and 100 ms, $ARDP$ values of LCC are roughly maintained at
values between 1.4 and 2 for large overlays. OMNI has almost a
constant $ARDP$ value (1.82) and performs on average better than
LCC in 12000-nodes overlay. We also note that in large overlays,
for clusters defined with 10 ms as node's scope, $ARDP$ increases
to reach 3, as nodes are more likely to be scattered. Larger
scopes are more efficient in this case.

\subsubsection{Stress}

\begin{figure} [htb] \centering
\includegraphics[width=3.4in]{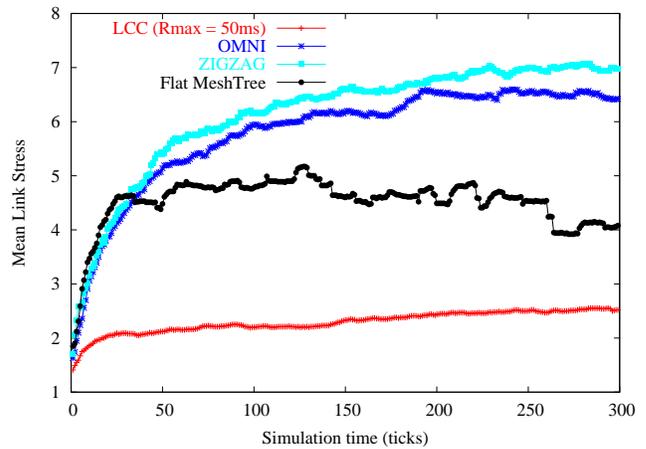}
\caption{Simulations: Stress on the links.} \label{StressvsTicks}
\end{figure}

Fig.~\ref{StressvsTicks}  shows average physical network stress for each of the overlays,
namely OMNI, ZigZag, Flat MeshTree and LCC ($R_{max}=50ms$) 2000
seconds after the last node joined. OMNI and ZIGZAG stress values
stabilize between 6.5 and 7. The Flat MeshTree overlay leads to
somewhat lower stress than OMNI and ZIGZAG with stress highly
oscillating between 4 and 5 due to random connections established
by newcomers. We note finally that LCC has an impressively low
stress, 2 to 3 times less than other overlays, with a steady
stress value between 2.5 and 2.8. Topology information is of
paramount importance in this observation, as packets sent through
the top-level hierarchy are sent to the cluster leader and in some
cases to potential edge nodes. Our clustering process allows then
to reduce the amount of redundant flows entering each network,
considering clusters as ``local networks'', and cluster leaders as
``multicast-enabled routers''.

\subsubsection{Robustness}

\begin{figure}
[htb] \centering
\includegraphics[width=3.4in]{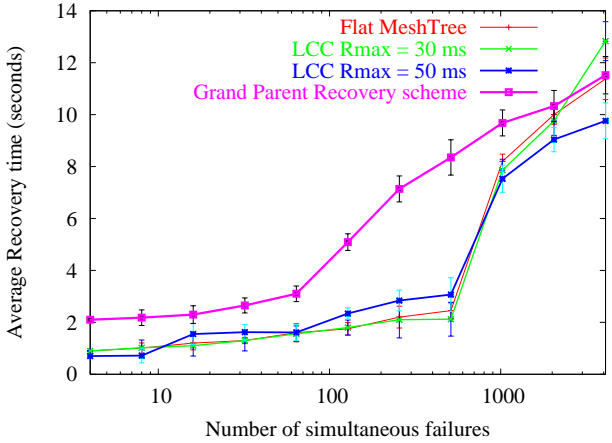}
\caption{Simulations: Failure Recovery time property.}
\label{FailureRecovery}
\end{figure}

When a non-leaf node quits, the overlay needs to be reconstructed.
So, it is important that this node's children can quickly locate a
new parent to resume the session. In addition, the recovery
process should not result in a fan-out violation in any node. In
LCC, to recover from the failure of its neighbor a node has to
redirect packet requests from that neighbor to another nearby in
its proper cluster. We compare LCC to the \emph{grandparent
recovery scheme} studied in~\cite{proactive}. In this scheme, the
children of the node which departs try first to attach to their
grandparent provided that the latter has enough capacity.
Otherwise, they are redirected to its descendants. We instruct a
number of randomly selected nodes in a 5000-nodes overlay to leave
the session simultaneously. Then, we observe the recovery time of
members, as the average time for an affected node to resume the
session. Results in Fig. \ref{FailureRecovery}  show that LCC
always yields a smaller recovery time than the tree-based
grandparent recovery scheme. On average, LCC takes 3.85 seconds to
recover from failures, which is about 35\% less than for the
grand-parent recovery scheme.

\begin{figure} [htb]
\centering
\includegraphics[width=3.4in]{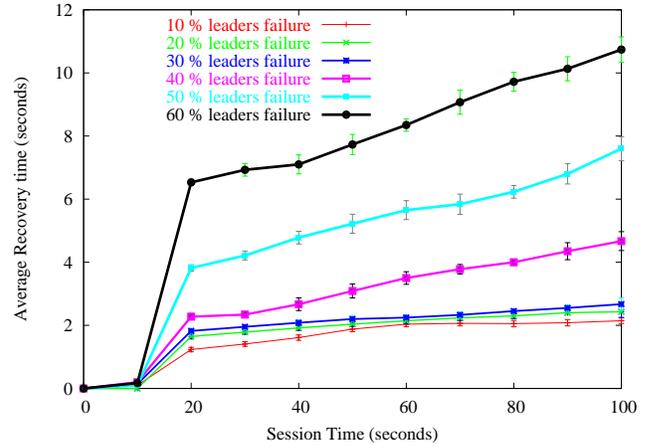}
\caption{Simulations: Impact of Leaders failures.}
\label{RecoveryLeaderFailure}
\end{figure}

In Fig.~\ref{RecoveryLeaderFailure},  we study the capacity of the LCC overlay to recover from cluster
leader failures. Each 10 seconds, a set of randomly selected
cluster leaders are instructed to simultaneously leave a
5000-nodes LCC overlay ($R_{max}$ = 50 ms). We observe that when
30\% of cluster leaders fail simultaneously, the recovery time is
almost $\leq$ 2 seconds. LCC is robust thanks to: 1) the proactive
rescue plan of leaders election and 2) the edge nodes connected to
the top-level topology, that allows to achieve data in case of
leaders' data disruption.

\subsubsection{The locating process efficiency}

\begin{figure} [htb] \centering
\includegraphics[width=3.4in]{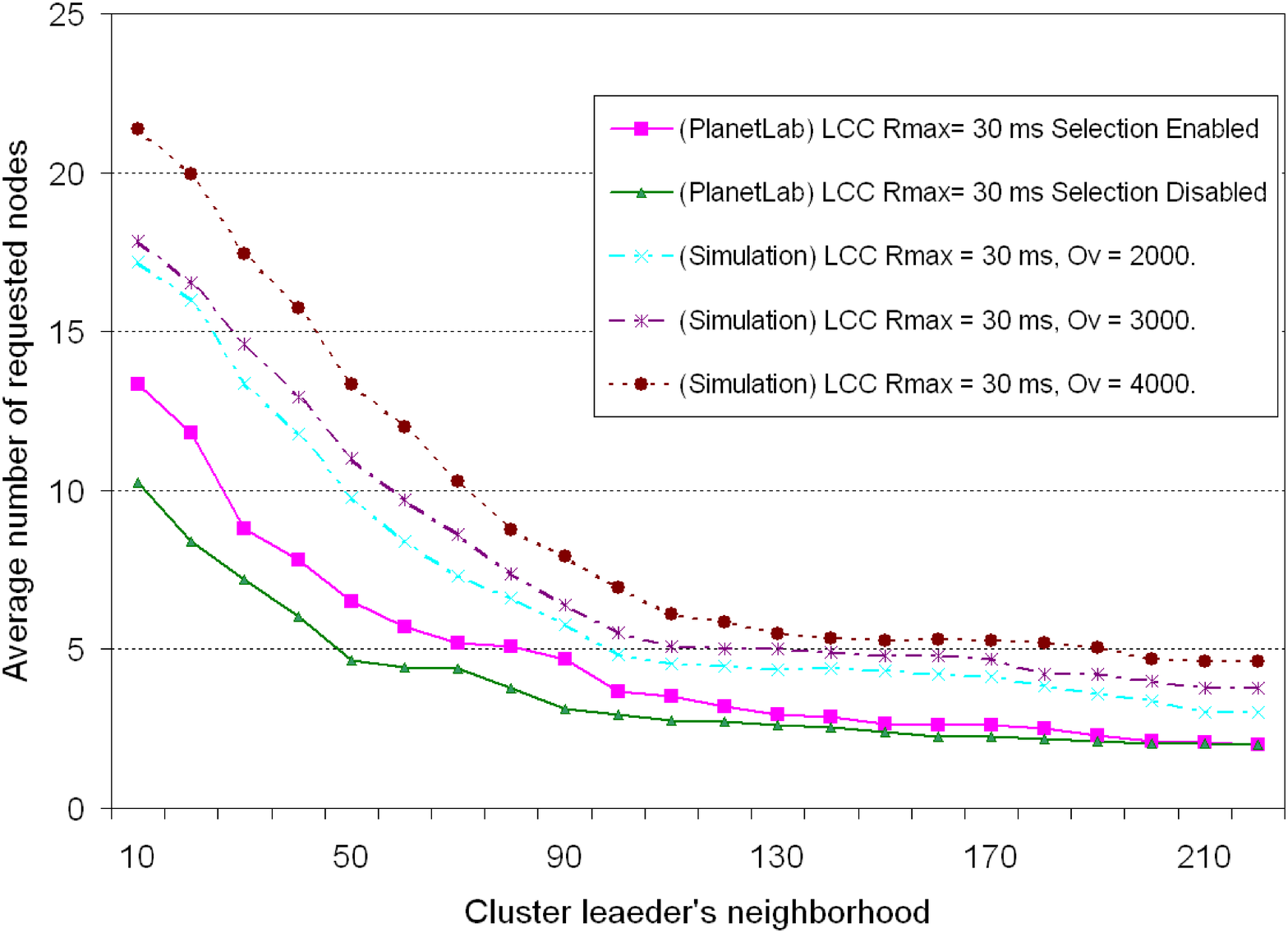}
\caption{Average number of requested nodes.}
\label{hopsKnownneighbors}
\end{figure}

To evaluate the behavior of newcomers during the locating process,
we observe the average number of requested nodes contacted by a
newcomer. Fig.~\ref{hopsKnownneighbors} depicts the average number
of requested nodes versus the total number of known cluster
leaders in each requested node's locating system. The figure plots
both PlanetLab results and simulations of 200 newcomers running
the selective locating process, once the overlay size reaches
respectively 2000, 3000 and 4000 nodes, resp. denoted by Ov =
2000, Ov = 3000 and Ov = 4000 in Fig.~\ref{hopsKnownneighbors}.


\begin{figure} [htb] \centering
\includegraphics[width=3.4in]{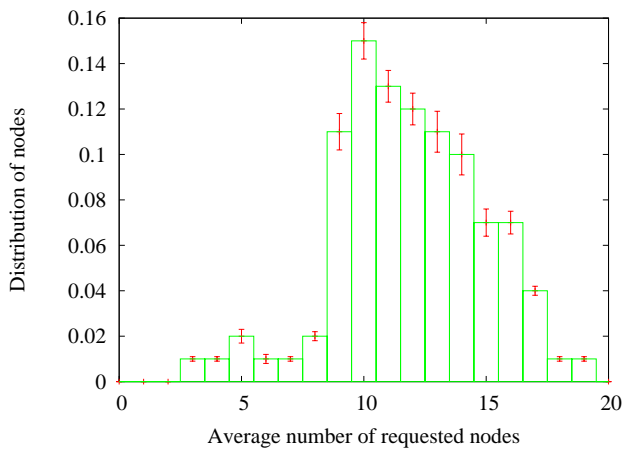}
\caption{Simulations: Distribution of nodes during the locating
process.} \label{DistributionAverageHops}
\end{figure}

We expected that the selective locating process needs more
requested nodes than the non selective process. Indeed, since it
queried representative nodes at each iteration, it may be less
accurate in one iteration, and hence requires to contact more
nodes afterwards. Results show however, that the selection has
little impact on the locating efficiency. The selective locating
process performs almost as well as the non-selective process, with
a maximum of 13 requested nodes in PlanetLab experimentations.
Moreover, curves are very close when the number of known cluster
leaders is large. We also observe that the locating process scales
well to large overlays: In a 4000-nodes overlay, newcomers are
located by contacting less than 12 requested nodes that know about
only 60 nodes in their locating system. The distribution of nodes
depicted in Fig. \ref{DistributionAverageHops} shows that more
than 80\% of a 2000-nodes overlay are able to terminate the
locating process by contacting less than 15 requested nodes. On
average the locating time in the experiment is very low with a
mean locating time of 3.2 seconds, a maximum of 7.2 seconds and
minimum of 1.8 seconds. Finally, we note that 98.4\% of newcomers
are able to connect to their closest cluster upon their arrival.
300 seconds after the last node joins the overlay, 99.3\% of nodes
are connected to their closest node. This demonstrates the
locating process accuracy, which is one of the reasons for the
resulting promising performances of LCC.

\section{Conclusion}

In this paper, we proposed a practical solution to enhance
different QoS aspects of overlays, namely scalability and
efficiency. The overlay construction is initiated by a simple and
scalable locating process that allows newcomers, after contacting
a few nodes, to locate the closest cluster in the overlay. The
locating process includes a selection algorithm to minimize
measurement overhead. On the basis of the locating process, we
proposed an hierarchical topology-aware overlay construction. We
introduced mechanisms to pro-actively deal with leaders failures
and underlying topology characteristics changes. Our PlanetLab and
simulations experiments prove that LCC incur low overhead during
both localization and data distribution. Compared to other
enhancement-based and topology-aware approaches, LCC achieves
shorter convergence time and performs less link adjustments rate.
At the same time, the scheme is robust to nodes' failures and
performs well in terms of data distribution efficiency especially
in large overlays. In conclusion, we believe that LCC is very
suitable for large-scale applications such as event broadcast for
thousands of attendees. In future works, we will focus on ways to
automatically tune different parameters such as nodes' scope and
stop criterion, through real-life tests. We will also investigate
techniques to secure the overlay and detect/prevent users from
cheating.

%
%


\begin{thebibliography}{18}

\bibitem{ESM}   Y.~H.~Chu, S.~G.~Rao, and H.~Zhang, \emph{A case for end system multicast}. In ACM
SIGMETRICS, Santa Clara, June 2000.

\bibitem{swithctrees}   D.~A.~Helder and S.~Jamin, \emph{End-host multicast communication using switch-trees
protocols}. In GP2PC, Berlin, May 2002.

\bibitem{Hostcast} Z.~Li and P.~Mohapatra, \emph{Hostcast: A new overlay multicast protocol}. In IEEE ICC, Anchorage (Alaska), June 2003.

\bibitem{MeshTree}  S.~W.~Tan, A.~G. ~Waters, and J.~Crawford, \emph{Meshtree: A Delay optimised Overlay Multicast Tree Building
Protocol}. Tech. Report 5-05, U. of Kent, April 2005.

\bibitem{OMNI}  S.~Banerjee, et al., \emph{Construction of an Efficient Overlay Multicast Infrastructure for
Real-time Applications}. In IEEE Infocom, San Francisco, March
2003.

\bibitem{LA05}  L.~Lao, et al., \emph{ TOMA: A Viable Solution for
Large-Scale Multicast Service Support}. In IFIP Networking,
Waterloo Ontario, May 2005.

\bibitem{ZigZag}    D.~Tran, K.~Hua, and T.~Do, \emph{Zigzag: An Efficient
Peer-to-Peer Scheme for Media Streaming}. In IEEE Infocom, San
Francisco, March 2003.

\bibitem{NICE} S.~Banerjee, B.~Bhattacharjee, and C.~Kommareddy, \emph{Scalable Application
Layer Multicast}. In ACM SIGCOMM, Pittsburgh, August 2002.

\bibitem{S004}  J.~K.~Sollins, \emph{Exploiting Autonomous System Information in Structured
Peer-to-Peer Networks}. In ICCCN, Chicago, October 2004.

\bibitem{KW02}  M.~Kwon and S.~Fahmy, \emph{Topology-aware overlay networks for group
communication}. In NOSSDAV, Miami Beach (Florida), May 2002.

\bibitem{Top-CAN}  S.~Ratnasamy, et al., \emph{Topologically-Aware Overlay Construction and Server
Selection}. In IEEE Infocom, New York, June 2002.

\bibitem{Meridian}  B.~Wong, A.~Slivkins and E.~G.~Sirer, \emph{A Lightweight Approach to Network Positioning
without Virtual Coordinates}. In ACM SIGCOMM, Philadelphia, August
2005.

\bibitem{PlanetLab}  \url{http://www.planetLab.org}

\bibitem {BRITE}   A.~Medina, et al., \emph{BRITE: Universal topology generation from a user's
perspective}. Tech. Report TR-2001-003, Boston, January 2001.

\bibitem{SA02} S.~Saroiu, P.~K. Gummadi, and S.~D.~Gribble,
\emph{A measurement study of peer-to-peer file sharing systems}.
In MMCN, San Jose (California), January 2002.

\bibitem{IMPLEM}    \url{http://www-sop.inria.fr/planete/software/LCC}

\bibitem{KEN04} K.~Shen, \emph{Structure Management for Scalable Overlay Service
Construction}. In USENIX NSDI, San Francisco, March 2004.

\bibitem {proactive} M.~Yang and Z.~Fei, \emph{proactive
approach to reconstructing overlay multicast trees}. In IEEE
Infocom, Honk kong, March 2004.

\end{thebibliography}
\end{document}